%% file: main.tex
\documentclass[sigconf,nonacm]{acmart}
\AtBeginDocument{%
  }

\usepackage{subcaption}
\usepackage{xspace}
\usepackage{tikz}
\usepackage[table]{xcolor}
\usepackage{pifont}
\usepackage{graphicx} % ensure this is in your preamble
\usepackage{adjustbox} % if more layout control is needed
\usepackage{multirow}

\include{defs}

\begin{abstract}
Unstructured data, such as text, images, audio, and video, comprises the vast majority of the world’s information, yet it remains poorly supported by traditional data systems that rely on structured formats for computation. We argue for a new paradigm, which we call computing on unstructured data, built around three stages: extraction of latent structure, transformation of this structure through data processing techniques, and projection back into unstructured formats. This bi-directional pipeline allows unstructured data to benefit from the analytical power of structured computation, while preserving the richness and accessibility of unstructured representations for human and AI consumption. We illustrate this paradigm through two use cases, and present the research components that need to be developed in a new data system called \sys. 

\end{abstract}

\begin{document}

\title{A Case for Computing on Unstructured Data}

%\title{\systemname: Enabling Computation Over Unstructured Data}
\author{Mushtari Sadia}
\affiliation{%
  \institution{University of Michigan}
  \city{Ann Arbor}
  \country{USA}}
\email{mushtari@umich.edu}

\author{Amrita Roy Chowdhury}
\affiliation{%
  \institution{University of Michigan}
  \city{Ann Arbor}
  \country{USA}}
\email{aroyc@umich.edu}

\author{Ang Chen}
\affiliation{%
  \institution{University of Michigan}
  \city{Ann Arbor}
  \country{USA}}
\email{chenang@umich.edu}

\maketitle

\input{sections/1_intro}

\input{sections/2_text}

\input{sections/3_diagrams}
\input{sections/4_system_overview}
\input{sections/5_experiments}

\input{sections/related}

\bibliographystyle{ACM-Reference-Format}
\bibliography{custom}
\end{document}

%% file: defs.tex
\newcommand{\sys}{\textsf{MXFlow}\xspace}
\newcommand{\neural}{ (\tiny{$\mathcal{N}$})}
\newcommand{\symbolic}{ (\tiny{$\mathcal{S}$})}

%% file: sections/1_intro.tex
\section{Introduction}
\label{sec:intro}

%\arc{WIP}

The majority of today’s data is unstructured---natural language text, images, audio, or video. This is not only because the world inherently produces analog signals, but also because the human brain is naturally adept at interpreting unstructured inputs. While our world and our cognition are fundamentally geared toward unstructured content,  our computing machinery excels at manipulating structured formats. 
This mismatch means that much of the world's data has been effectively locked away from our computing tools. 
Despite decades of progress, data systems require that the data already adheres to a structured format (e.g., tables, graphs, programs). 
%When a usable structure is not readily available in the data, traditional systems have remained largely ineffective. 

Recent advances in AI, particularly large language models (LLMs) and vision models, are beginning to shift this landscape. A key insight is that real-world data is "unstructured" only in its format—not in its meaning. Such data often contains rich, latent semantic structure that is intuitive to humans. AI models, like humans, can interpret and reason over this unstructured data and, when guided appropriately, can extract and articulate its underlying structure.

Capitalizing on this capability, a recent body of work has demonstrated promising results in analyzing unstructured data%including semi-structured document understanding, querying unstructured text with SQL, video analytics, and more
~\cite{lin2024towards, wang2025andbbreakingboundariesainative, biswal2024text2sqlenoughunifyingai}. Notably, Madden et al.~\cite{sammadden} envision a future where we can "query all of the world’s bytes" using AI. However, existing work has largely focused on one-way translation: converting unstructured data into structured formats to just enable querying.

In this paper, we argue for a broader and more ambitious vision: a new paradigm of \textit{computing on unstructured data}. Our vision goes \textit{beyond} querying—it involves computing in the full sense of the term. Specifically, querying is essentially a read-only operation that allows efficient information retrieval. In contrast, by \textit{computing} we refer to a read/write paradigm where one can manipulate, interact with and reason over unstructured content. 
To operationalize this vision, we propose a holistic eXtract–Transform–Project (XTP) pipeline.  The \textit{extract} phase ingests raw unstructured data, and converts it into a structured internal representation- for instance, transforming text into tables or images into symbolic programs, or audio into event timelines. The \textit{transform} phase applies data processing techniques (e.g., optimization, imputation, validation) over this structured representation. Finally, the \textit{project} phase translates the transformed results back into unstructured formats, either the same as the original or a different one, making the outputs easier for humans and AI agents to consume and act upon.

A central feature of the XTP pipeline is its \textit{bidirectional} nature. 
While prior work has looked at each of these phases individually, XTP closes the loop and enables end-to-end computation. This mirrors how humans think: we extract latent structure to reason, then re-embed those insights in unstructured ways (e.g.,language, visuals, audio) to communicate. The intermediate structured representation serves as a computational substrate, enabling principled and verifiable transformations, while the final projection step ensures that results remain compatible with how data is consumed in practice. By tightly integrating structure and unstructured representation in a single loop, XTP unlocks the best of both worlds: it retains the expressiveness and richness of unstructured data, while unlocking the analytical power and rigor of structured computation.

Consider two motivating examples. \textit{(i) Text:}    Doctors often work with free-form text, such as medical prescriptions. An XTP pipeline can automatically \textit{extract} patient information and diagnoses to build a structured database. While querying this data (e.g., “Which patients were prescribed Drug X?”) enables retrieval, real-world tasks often require more, such as adjusting dosages or identifying at-risk patients based on complex factors. These require direct computation and data modification.  The \textit{transform} phase enables such updates, and the \textit{project} phase generates human-readable outputs like revised prescriptions or reports.
\textit{ (ii) Diagrams}:
Engineers use complex visual artifacts (e.g., datacenter layouts, schematics, blueprints) rich in information but hard to process. XTP can \textit{extract} formal representations (e.g., a symbolic program showing the components and connections) of the visual artifacts, whether circuit schematics or datacenter layouts. 
While simple queries (e.g., “Does this node have upstream power?”) aid inspection, tasks like validating redundancy or optimizing fault tolerance require transform operations: editing components, simulating
flow, or refactoring layouts. The \textit{project} phase regenerates an updated visual diagram for engineers to inspect and refine.
\begin{comment}\textit{(ii) Diagrams}:   
Engineers frequently work with visual artifacts like datacenter layouts, circuit schematics, or architectural blueprints. We could convert diagrams into programs that encode components and interconnections in a machine-readable format. %into a graph where nodes represent components and edges capture electrical connections.
This further enables validation (e.g., datacenter redundancy), reasoning about upgrades (e.g., from N+1 to 2N redundancy), which not only require data querying but also graph manipulation that modify components and circuits. Updated programs can then be projected back into a visual diagram that engineers can inspect, annotate, and revise.
\end{comment}
%In both of these cases, the end-to-end XTP loop unleashes opportunities for programmatic transformations that would otherwise be impossible on unstructured data.  

\begin{figure*}
    \centering
    \includegraphics[width=0.9\linewidth]{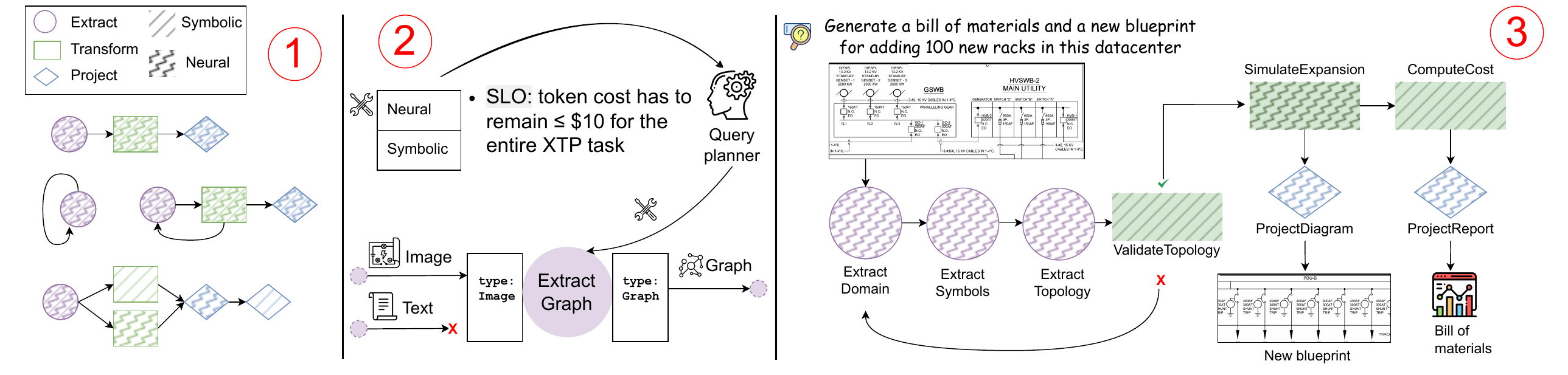}
   \vspace{-0.3cm} \caption{A sketch of our envisioned design. \textcolor{red}{\textcircled{1}} Shows operator patterns, where operators can be neural, symbolic, or hybrid, with possible self-loops or dynamic routing. \textcolor{red}{\textcircled{2}} Depicts an operator for which the query planner selects neural or symbolic tools based on criteria such as service-level objectives (SLOs). These objectives apply to the full end-to-end task and must be considered holistically during tool selection. Operators (e.g., \texttt{ExtractGraph}) consume data of enforced types (e.g., \texttt{type: Image}) and produce outputs in specific formats. \textcolor{red}{\textcircled{3}} Shows an example query, where the query planner orchestrates a concrete sequence of XTP operators to extract structure, apply transformations, and project an updated diagram and bill of materials report.}

    \label{fig:e2e}
\end{figure*}

Achieving this vision requires addressing a range of research challenges. Unstructured data is inherently more ambiguous and fuzzy than what conventional data systems can deal with. A na\"{i}ve approach is to ``throw everything at the LLM,'' but LLMs are non-deterministic, prone to hallucination, and often make subtle errors. Additionally, they are difficult to integrate as computing modules due to their unpredictability and variability in performance, cost, and failure modes. 
We argue that tackling these challenges requires rethinking data system design, specifically, through a new system specialized for unstructured data management, which we call \sys. \sys allows users to express tasks as a dataflow graph, where nodes represent operators and edges encode the flow of data. This structure supports a principled decomposition of tasks into subtasks, each of which can be independently scheduled, executed, and validated using tools tailored to the specific operation. Yet, \sys departs from traditional dataflow systems in several ways. 

The first is the use of \textit{hybrid tools.} \sys supports both neural and symbolic operators. While large language and vision models offer powerful tools for handling unstructured inputs, they need to be coupled with symbolic methods that are deterministic and come with strong correctness guarantees. \sys enables seamless composition of these tool types—leveraging the flexibility and expressiveness of neural models when needed, while falling back on symbolic techniques, built on decades of data management research, wherever possible. 
%\item \textbf{Flexible operator design.} Processing unstructured data woukd entail desigining novel operators. In particular, itransporm ohase, working on mutil modal data.  The best structural representation could vary from dataset to dataset, or even from record to record.  
%best representation of the structure itself  often requires novel operator types not found in conventional systems. This necessitates new methods for query planning—translating logical workflows into physical execution plans that balance cost, performance, and accuracy. Additionally, since users may not be able to fully specify their intent up front (e.g., the best transformation method might depend on unseen input data), \sys must support a unified and extensible framework that allows users to define and compose custom operators interactively.
%Processing unstructured data often requires custom operators that span multiple modalities or data representations. Since structure can vary from one dataset—or even one record—to another, \sys supports a flexible operator framework, allowing users to build and compose novel transformations with minimal boilerplate.
Moreover, \sys enables \textit{dynamic dataflow.} Unlike traditional dataflow systems with fixed control paths, \sys supports dynamic routing based on the data itself. Depending on content characteristics or intermediate results, the system may direct data to different downstream operators or broadcast it to multiple paths for parallel processing. This flexibility allows \sys to adapt to the heterogeneity of unstructured inputs and the uncertainty inherent in their processing.

%% file: sections/2_text.tex
\vspace{-2mm}\section{A Deeper Dive on Motivating Examples}
\vspace{-1mm}

Our goal is to enable classes of computation on unstructured data: 

\textit{(i) Data protection:} Structure enables precise, transformation-based privacy protections that querying alone cannot provide. For example, differential privacy can be applied to structured aggregates, synthetic datasets can be generated for safe sharing, and role-based access control can be enforced through filtered views of the structured data. These protections require direct manipulation of the data representation—again going beyond what is possible through querying, which cannot reshape or sanitize the data itself.

\textit{(ii) Data provenance:} Despite much progress in data provenance and its widely-known utility (e.g., auditability, reproducibility, explainability), tracking provenance over raw unstructured content is challenging—text, images, and other media rarely carry explicit links between inputs and outputs, making it hard to reason about how a result was derived. Structured transformations overcome this by enabling detailed records of how each extracted element was generated, identifying which parts of the original input contributed to it, and tracking its evolution through subsequent steps.

\textit{(iii) Data cleaning:} Cleaning raw unstructured inputs directly is difficult. By extracting structured representations, XTP can reveal inconsistencies in the raw input—e.g., mismatched attribute values, redundant entities, or violations of an implicit schema. Once in structured form, these issues are easier to detect and resolve, unleashing existing or new data cleaning techniques, such as imputing missing values, deduplicating across heterogeneous sources, and normalizing inconsistent representations (e.g., units, dates, or names), before projecting the data back to unstructured domains.

In short, a range of transformations become possible to improve the quality or usability of unstructured data.

We now dive into two examples as further illustration, describing the end-to-end pipeline with challenges and potential ideas. 
%. These include schema refactoring to improve compactness, converting data into alternative logical representations, or simplifying and optimizing computational models for enhanced performance and readability. Such transformations are essential to convert unstructured content into higher-quality, more manageable data.

 %A large portion of the world's knowledge is stored in free-form text, which is not directly compatible with structured data systems such as relational databases. Despite a long line of work in Information Extraction (IE)~\cite{Mansuri-uns-to-db,dmg1,dmg2,dmg3,dmg4,dmg5,dmg6}, these brittle, rule-based approaches are quickly outperformed by recent work powered by large language models (LLMs), which can convert text into database tables~\cite{Aroravldb, wu2022texttotable, sundar2024gtbls, li2023seq2seqset} to support SQL queries. \sys pushes this vision much further beyond ease of querying.
 \vspace{-3mm}
\subsection{Computing on unstructured text} 
 
Consider medical text, such as physicians' notes, admission forms, and prescriptions from clinical trials, which contains valuable information that can be shared with other hospitals, medical analysts, healthcare policy makers, or used to train medical NLP models. %However, the text may contain sensitive information (e.g., patient's diagnosis, locations), it may be low-quality due to hasty writing (e.g., misspellings), and could be noisy (e.g., different drug brand names may correspond to the same generic drug). Physicians may make intentional or accidental changes to previous notes, which need to be tracked and analyzed. %Although well-established techniques (e.g., data provenance, differential privacy) exist, they work most effectively over relational databases and are handicapped by the raw text format. \sys would enable such computation via the EXP pipeline: extracting a relational database from the notes, transforming the database records for deduplication, privacy, and provenance, and finally projecting the results back to a synthesized textual data that can be shared with a broader audience.

%\sys would help extract a relational database from the prescriptions, deduplicate entries (e.g resolving multiple mentions of the same patient, drug names such as mapping different brand names to their generic drug names), apply differential privacy mechanisms to aggregate queries (e.g finding the most prescribed antibiotics for pneumonia while protecting individual treatment histories), and finally, it would project the result back to a synthesized textual dataset that can be shared with a broader audience without leaking any sensitive information. 

\noindent \textbf{Extract.}
The first step is to \textit{extract} a relational database from the prescriptions.  
\sys must first infer a database schema that captures key entities and relationships. Unlike prior work in Text2Table or Text2SQL~\cite{deng2024text}, most existing methods either extract only flat tables, assume a predefined schema, or focus solely on translating queries. Thus, automatic schema generation from raw text remains a critical challenge, one that we took the first step in addressing in recent work~\cite{squid}, by defining the task of synthesizing relational databases from unstructured text. 
Once a suitable schema is generated, the next task is to populate the database with records from the text. This involves (i) extracting all relevant values comprehensively, (ii) ensuring consistent representation of entities across tables using shared identifiers such as primary and foreign keys, and (iii) assigning each value to its correct column.
Finally, \sys needs to construct a database instance using SQL statements, such as \texttt{CREATE TABLE} and \texttt{INSERT INTO}.

\noindent\textbf{Transform.} 
\sys can now apply various transformations on this database.
Consider data cleaning: errors from the original text would be reflected in the extracted database. Using the relational schema, \sys can address these issues in a principled way. For example, it may detect missing values in certain columns (e.g., timestamps) and extrapolate them from neighboring records. It can impute data when patient attributes are incomplete but can be inferred from similar cases (e.g., patients with comparable age and symptoms). \sys can also repair inconsistencies, such as merging multiple brand names under a common generic drug or standardizing dosage formats.
As another example, consider data protection:
if the data is to be shared with a broader audience, \sys allows specifying transformations to ensure privacy. For instance, differential privacy can be used to release answers to statistical queries in a privacy-preserving way. Alternatively, one can anonymize the data by dropping columns like "name" or "SSN" using a SQL query, which is not easily done when sensitive content is mixed with unrelated information.

%\textbf{Data provenance.} Provenance tracking also becomes easier at the tuple or table level. Each row can carry metadata that records where it came from: the document ID, the exact line number, and the sequence of transformations it underwent (e.g., normalization, deduplication, schema alignment). This opens up new capabilities for audit and accountability---e.g., a medical analyst discovering an unusually high dosage can trace that value back to the specific physician note. Several interesting challenges include: a) a single sentence in the source text may give rise to multiple structured rows, or vice versa, making one-to-one mappings insufficient; b) transformations that clean or modify the database (e.g., merges duplicate records or imputes missing values) also need to be tracked, together with the lineage of unstructured data extraction, to attribute changes to specific portions of the original input. Hence, \sys provenance support must extend classic database provenance~\cite{wah-chieh-whyandwhere, peter-buneman-papers} to address these issues. 

\noindent\textbf{Project.}
Finally, cleaned prescription data can be rewritten as discharge summaries, treatment explanations, or visualizations (e.g., charts), therefore, \sys needs to support output across multiple modalities. The first step is \textit{modality selection}, choosing the best representation for a given table or column. For instance, antibiotic dosage data may be projected as a chart, pneumonia treatment plans as text, and diagnostic results as synthetic chest X-rays. A single discharge report may combine several of these formats, requiring careful cross-modal alignment to ensure coherence.
Once modalities are selected, \sys generates outputs that integrate both the structured database and original raw text. This involves producing content that is not only factually accurate but also contextually appropriate.

%% file: sections/3_diagrams.tex
\vspace{-0.1cm}\subsection{Computing on engineering diagrams}

% \ms{ready for a pass.}
%Engineering diagrams are essential for designing complex systems like power grids, HVAC layouts, and datacenter infrastructure—where multiple diagrams (e.g., network topology, rack layout, power distribution) must align. Today, experts manually inspect these diagrams for faults and optimizations, a process that is slow and error-prone. Prior work has focused on low-level tasks like symbol detection and connection tracing, but \sys will enable systematic computation far beyond existing work. 

Consider a datacenter blueprint drawn to U.S. electrical codes, which needs to be converted to meet Canadian standards. This involves adjusting units (e.g., wire gauge, voltage), enforcing country-specific grounding and routing rules, resolving layout issues, and generating updated blueprints or compliance documentation~\cite{qrfs2025uscacodes}. 
While prior work has focused on low-level tasks like symbol detection and connection tracing (i.e., ``queries''), \sys will enable end-to-end computation for the above transformation task.

\noindent \textbf{Extract.} 
A structured representation of the datacenter blueprint could be viewed as a program that describes its topology. The first step is to perform \textit{domain classification}, identifying the diagram type: datacenter blueprint, HVAC layout, or network schematic? This is similar to schema generation in databases: only after the domain is known, the system can proceed to interpret the symbols and structure. Example steps are: (i) diagrams from different domains can look visually similar (e.g., electrical circuit diagrams vs. digital logic designs); (ii) some diagrams require fine-grained classification---e.g., distinguishing between single-line and a three-line electrical diagrams. %(3) The classification process can be recursive; symbol resolution may update the original domain guess. %To address these challenges, \sys~can first guess the domain using a multimodal language model, then refine it via retrieval-augmented generation (RAG) by retrieving symbol libraries specific to that domain. The initial classification can be verified after symbol resolution.
The next \textit{symbol resolution} step is to map visual symbols to their semantic identities. This requires disambiguating similar shapes that mean different things across domains (e.g., a triangle might represent an amplifier in analog circuit diagrams but a multiplexer in digital logic); and reconciling symbol variations due to differing standards (e.g., resistor shapes can vary between IEC and ANSI standards,  metric vs. imperial system). 
%Symbol ambiguity can arise in several ways: (1) conventions may vary across domains (e.g., a triangle might represent an amplifier in analog circuit diagrams but a multiplexer in digital logic); (2) even within a domain, symbol sets may differ by vendor or tool (e.g., resistor shapes can vary between IEC and ANSI standards); (3) symbols can differ across countries, including units and labeling styles (e.g., metric vs. imperial). In such cases, ambiguous symbols can be reconciled by identifying the closest match based on embedding similarity to known examples.
The \textit{topology construction} step infers their connections and constructs a structured connectivity graph. This includes detecting and tracing connectors—such as wires or pipes—even when they are partially occluded, or noisy;
and interpreting these connections according to different domain conventions—e.g., a wire crossing might indicate a junction in electrical diagrams but not in HVAC schematics. %\arc{why is this the case not clear, maybe replace with a different challenge} and (3) the inferred topology may revise the original domain classification. To handle these challenges, \sys can apply edge-tracing algorithms alongside learned models like graph neural networks \cite{gnn-aclp} to predict likely connections. It can also use symbolic post-processing to validate the graph against domain-specific constraints. The resulting topology can then be converted into a structured format such as a netlist or infrastructure description language. 
%\arc{we talked about graphs before now we are talking about netlist/IDL}

% \ms{Domain classification and symbol resolution needs to be split into two different paragraphs. In domain classification, novel challenges need to be discussed, similar to different challenges discussed in symbol resolution, but challenges that are unique to domain classification. For example, does domain classification just mean identifying datacenter blueprint vs HVAC layout vs network diagram etc, or does it mean identifying further, such as what type of network diagram exactly? Also, is this a recursive problem, such as after resolution of symbols, our decision of the classified domain can change? Also, for domain classification, the analogy to schema generation needs to be drawn explicitly.}

% \ms{Challenges need to be articulated for constructing the topology. For example, inferring connections between components, what type of techniques are required here (e.g., GNNs)? What are some unique challenges (e.g., different conventions for cross-wire connections?) Can the identified connections or topology require rethinking of the domain classification? Is this also a recursive decision?}

\noindent \textbf{Transform.}
\textit{Design validation} is an example transformation. \sys can analyze the program and automatically check for correctness against domain-specific rules: e.g., verifying that only valid port-to-port connections exist (e.g., power source connected to load, not load-to-load), that all units are normalized (e.g., converting amperes and volts to a consistent base), or that the resulting topology satisfies required properties (e.g., tree vs. mesh connectivity). %These validation steps help catch design errors early and ensure that diagrams conform to operational or regulatory standards.
\textit{Design optimizations} are also made possible, 
where \sys improves or adapts the design programmatically. For instance, adding redundant components to a datacenter blueprint to increase fault tolerance (e.g., increasing the redundancy level from N+1, which tolerates one failure, to 2N, where each component has a standby replica); converting between region-specific conventions (e.g., U.S. to Canadian standards); finding topological alternatives that achieve the same functionality with improved energy efficiency. %(e.g., upgrading a datacenter to Tier III); reconfiguring the layout to accommodate datacenter expansions (e.g., adding 100 new racks).

%\arc{No need for provenance here}
\begin{comment}\textbf{Tracking provenance.}
After each transformation, \sys will record the sequence of changes to the structured diagram, enabling full auditability and rollback. Every connection or component can carry metadata describing its origin, such as the transformations applied on it. This is particularly valuable when multiple diagrams are merged or when optimization steps alter the design: engineers can trace each element's history, understand how decisions were made, and revert to earlier versions if needed.
\end{comment}

\noindent \textbf{Project.}
The modified programs are then projected back into human-readable formats, such as diagrams (e.g., revised designs) or textual artifacts (e.g., manuals or cost estimation reports). Generating diagrams differs from text generation, as it requires reasoning about spatial layout. This process can be decomposed with the help of intermediate representations, such as TikZ code.
\vspace{-0.1cm}
\subsection{Shared Challenges}
\label{subsec:shared-challenges}
Although we have discussed text and diagrams separately, they are both instances of the same larger problem: real-world scenarios often involve multimodal unstructured data that must be processed holistically. They
expose a common set of challenges for all three phases of the XTP pipeline. While extraction, transformation, and projection have each been studied independently, and some existing methods can plug into \sys as modular operators, isolating them misses the point. Unique challenges emerge when the XTP pipeline is viewed as a bidirectional, end-to-end system, where changes in one phase must propagate coherently across others. 

% Although we have discussed text and diagrams separately, they are instances of the same larger problem. Real-world scenarios often involve multi-modal unstructured data: e.g., medical diagnosis may involve textual prescriptions and X-ray images; engineering design requires using both diagrams and manuals. Hence, the XTP pipeline needs to flexibly handle and convert across modalities. 
% Although the two use cases differ in both format and task specificity, they expose a common set of challenges spanning all three phases of the pipeline. While extraction, transformation, and projection have each been studied independently, and some existing methods can plug into \sys as modular operators, isolating them misses the point. Unique challenges emerge when the XTP pipeline is viewed as a bidirectional, end-to-end system, where changes in one phase must propagate coherently across others. This shift
% demands new methods that unify all three stages—driving the core
% vision behind \sys. 
\noindent\textbf{Extract.} Recall that in XTP, the intermediate structure serves only as a computational substrate that must support principled transformation while also enabling faithful projection back into unstructured form. As a result, no one-size-fits-all representation for a given input format (text, images, etc) exist; the best structure is application-dependent and often not known a priori. For example, text can be represented with multiple relational schemas. Traditional schema evaluation methods like normalization fall short here because the schema must also enable reverse projection into natural language or other modalities—demanding new metrics that balance fidelity, and reversibility. For engineering diagrams, extraction involves constructing a graph-like representation of components and their connections, while handling diverse diagram types (HVAC, power, networking) with unique symbolic conventions, layouts, and domain-specific syntax. Choosing the right intermediate format, such as a typed, attributed graph with domain-specific annotations, demands anticipating downstream needs. Extraction must therefore be adaptive and context-aware, not just a low-level parsing step.

% Recall that in XTP, the intermediate structure serves only as a computational substrate that must support principled transformation while also enabling faithful projection back into unstructured form. As a result, no one-size-fits-all representation for a given input format (text, images, etc) exist; the best structure is application-dependent and often not known a priori. For example, text can be represented with multiple relational schemas. Traditional schema evaluation methods like normalization fall short here because the schema must also enable reverse projection into natural language or other modalities—demanding new metrics that balance fidelity, expressiveness, and reversibility. For engineering diagrams, extraction involves constructing a graph-like representation of components and their connections, while handling diverse diagram types (HVAC, power, networking) with unique symbolic conventions, layouts, and domain-specific syntax. Choosing the right intermediate format, such as a typed, attributed graph with domain-specific annotations, demands anticipating downstream needs. Extraction must therefore be adaptive and context-aware, not just a low-level parsing step.\\
\noindent\textbf{Transform.} In standalone transformation tasks, it is common to treat the structured output in isolation, often disregarding how the structure was originally generated. However, in our setting, this approach falls short—tracking exactly how and which parts of the raw input contributed to the structured representation is essential. This challenge appears in both use cases. In textual inputs, for instance, a single user may give rise to multiple records in the structured database. Without this mapping, it is difficult to determine whether an individual was counted once or multiple times in queries like “patients treated for pneumonia”—this is critical in contexts such as differential privacy, where sensitivity analysis depends on bounding each person’s contribution to the output~\cite{wilson2019differentially}. In diagrams, the challenge manifests differently. Identical-looking components, such as multiple 10$\Omega$ resistors may be mistaken to be the same entity.  To prevent this, \sys must maintain a mapping between each component’s position in the original diagram and its corresponding entity in the structure. 
% This ensures that each distinct instance is treated independently during transformation and avoids semantic errors in the resulting representation.
%a single sentence in the source text may give rise to multiple structured rows, or vice versa, making one-to-one mappings insufficient; b) transformations that clean or modify the database (e.g., merges duplicate records or imputes missing values) also need to be tracked, together with the lineage of unstructured data extraction, to attribute changes to specific portions of the original input. Hence, \sys provenance support must extend classic database provenance~\cite{wah-chieh-whyandwhere, peter-buneman-papers} to address these issues. 
\\\noindent\textbf{Project.} The project phase in the XTP pipeline faces a distinctive and complex challenge: unlike standalone scenarios, it must reconcile and balance \textit{two} intertwined sources of content—the original unstructured input and the transformed structured data. For instance, when generating a summary report from text, \sys must preserve the improvements from data cleaning, while simultaneously maintaining the original document’s style. In diagrams, the situation can be even more intricate. Often, only parts of a diagram are optimized or transformed (e.g., symbols that differ between U.S. and Canadian standards), while the rest remains unchanged. The projection must faithfully reflect these targeted modifications, while preserving untouched sections. 

\vspace{-2mm}
\section{\sys System Sketch} 
In this section, we formulate the design principles motivated by the challenges outlined in Sec~\ref{subsec:shared-challenges}, and sketch the \sys architecture.
\vspace{-0.4cm} 
\subsection{Design Principles}
At the heart of \sys\ is a rethinking of unstructured data computation—not as isolated tasks, but as an integrated pipeline of extraction, transformation, and projection. With \sys, we want to able to achieve the following overarching goals:

\noindent\textbf{End-to-end processing.} XTP tasks must be guided by downstream use—e.g., whether extracted content becomes a CSV, or time series. Rather than stitching separate systems, \sys treats the entire pipeline as a single query driven by service-level objectives (SLOs).

\noindent\textbf{Multi-modality computation.} Each pipeline stage must flexibly operate across different modalities. For instance, a transformation step might flatten tables to prompt an LLM, or align symbols with textual descriptions. Projections must also choose or mix output modalities—e.g., generating text, or charts—based on task requirements, while preserving cross-modal consistency.

\noindent\textbf{Under-specification.} 
 Unlike traditional structured data systems, analysts often cannot fully specify queries upfront. Therefor, \sys must be dynamically adaptable: it should respond to shifts in data distribution or to unexpected model behaviors.

To realize these goals, \sys is built on the following core principles: (i) mix neural and symbolic operations, (ii) dynamically determine execution order, including recursive calls, and (iii) handle multi-modal data. These capabilities go beyond what traditional query interfaces can support. As a result, a \textit{dataflow}-based architecture is better suited to \sys's needs (Fig.~\ref{fig:e2e}). 
\\To illustrate how these principles translate to concrete capabilities, we walk through the stage-specific challenges identified in Sec \ref{subsec:shared-challenges}:
\\\textit{Extract.} \sys~addresses the challenge of not knowing the most suitable structure apriori by leveraging its  dynamic dataflow. This enables the incremental construction of structure, decomposed into a sequence of stages (e.g. schema generation, value extraction). At each stage, the output is routed through multiple candidate branches, allowing the system to evaluate and select the most suitable structure downstream. These intermediate results are verified using a combination of neural and symbolic tools: neural models can generate an initial schema or classify a diagram's domain, while symbolic methods enforce referential integrity, or check domain-specific constraints (e.g., component connectivity in diagrams).
\\\textit{Transform.} For the \textit{transform} phase, maintaining entity mappings with the original source is possible because \sys's dataflow model explicitly tracks each transformation step. As data flows through the pipeline, entity mappings can be updated and encoded at each node. In case of errors, the system can roll back to a previous node, apply a fix to a specific transformation, and then re-propagate the corrected result through the downstream stages.
\\\textit{Project.} For the \textit{project} stage, \sys uses neural models to generate content from structured data, while maintaining faithfulness to the original source. To prevent content that was intentionally removed during transformation from reappearing, especially when incorporating elements from the original unstructured input, \sys tracks all deleted values and applies symbolic checks to ensure they do not resurface in the output. We next show how these design principles are realized in the key building blocks of \sys.

%% file: sections/4_system_overview.tex
\vspace{-0.4cm}
\subsection{Data Model} 
\vspace{-0.1cm}
\sys's needs to process unstructured, semi-structured, and structured data in the same programming framework. Hence, its data model represents a departure from traditional data systems (e.g., MapReduce, Spark, Naiad), which expect a fixed and predefined data type (e.g., textual records, or streaming data tuples). 
We propose that \sys data should be \textit{typed}, 
with type refinement, conversions, and subtyping built into the system. This is inspired by recent work~\cite{ousterhout2023zed} that made a case for explicitly typing data fields in heterogeneous documents, but \sys further envisions automatically identifying types from data and using them to model transformations. 
For instance, \sys queries will consume data as \texttt{type Text} or \texttt{type Image}, and gradually transform these types into 
\texttt{type Table}, \texttt{type Graph}. Types such as 
\texttt{type Graph} can be further subtyped, e.g., as electrical diagrams in the US or Canadian convention, or converted explicitly or implicitly (i.e., coercions).   
Data types may also be generic (e.g., \texttt{type Unstructured}), if the underlying data represents a mix of multiple types or is simply unknown at compile time. In these cases, types may be assigned at runtime after structures have been detected from the underlying data. 
This not only helps \sys to handle flexible data processing, but also enables better composition across operators and queries---e.g., one \sys query can resume the work of another if their interfaces type check. 
 % \ms{add an example multiple types} \ms{hierarchy}

\input{tables/table1}
\input{tables/table2}
\subsection{Dataflow Queries}

Operators in a \sys query have typed input/output interfaces. For instance, \textit{extraction} operators consume unstructured data types, and transport them into the structured domain. Likewise, \textit{projection} operators achieve the reverse, and \textit{transformation} operators consume and produce typed data schemas. Furthermore, \sys operators may be neural (i.e., invoking an AI model) or symbolic (i.e., invoking a procedure); and they may be \textit{preprogrammed}, where the data analyst specifies the AI models or functions to invoke, or they could be \textit{fungible}, where operator functions are partially specified, or only described in prompts; in the latter case, the implementation is left to the query synthesis stage for ``auto-completion.''

Operators are connected via edges, forming a directed dataflow graph. However, unlike traditional dataflow systems, \sys edges could be probabilistic; they could even be late-bound, where the specific processing order is determined at runtime, and may further change over the lifetime of a single query, e.g., depending on whether sufficient structure was detected or whether additional LLM invocations are needed. For instance, an \texttt{ExtractGraph} operator may reroute outputs back to itself if the confidence is below 80\%, to reprocess the data using a more powerful LLM.

Despite this dynamic execution order, edges can only compose operators where the input/output types are compatible (possibly via implicit type coercions)---where the type system provides symbolic guardrails. 
For instance, if a transform operator expects a graph but receives a database, processing will be rerouted to another transform operator; if corresponding operators cannot be found, control flow may be rerouted back to the extraction phase, with an enforced graph output. 
Combined, these dataflow designs make \sys suitable for composing, optimizing, and executing complex data-centric workflows over mixed-modal data.

%\ms{should we add some examples here? e.g., if an extraction operator’s confidence is below 80\% it's rerouted to itself with a more powerful choice of model?}

%\ms{How would type coercions work? If e.g the transform operator is expecting a graph but receives a database, the pipeline is rerouted to another transform operator. If a corresponding operator is not found, it would be rerouted to the extraction phase, with an enforced graph output? Should we provide a concrete example, or keep it open-ended?}

\subsection{Query planning} 
\label{subsec:planning}

%-- autocompletion of operators, 
%-- generating the query itself from prompts. 

An \sys query encodes a logical plan, and the query planner is responsible for generating the best physical plan for execution. \sys's query planning stage involves several new challenges not found in traditional SQL and dataflow systems. 

To start with, a \sys query may be partially-specified. 
For fungible operators, the query planner may choose between using a rule-based extractor, a powerful LLM, or a locally-hosted model. For instance, an \texttt{ExtractField} operator could be implemented using CoreNLP libraries or AI methods. These implementations differ in token usage, execution time, and accuracy; and dataflow edges can also be chosen dynamically. We propose to cast this as a program synthesis problem, which takes a partially-specified query and auto-completes it to satisfy the intent while considering some service-level objective (SLO): e.g., 90\%+ accuracy for 95\% of the data. 
The decisions may vary across different parts of the dataset, as the SLOs are defined for the entire end-to-end task and must be considered holistically when selecting tools for each operator.

% \ms{multimodal data, open-ended tasks, no predefined operations, high level abstract map}
% The query itself could also be synthesized from prompts using a language model, allowing analysts to describe the desired operation in natural language. The system can then decompose this into a concrete operator graph by inferring the required operations, and then fill in the appropriate tools. Same as before, the planner selects specific tools for the operations by evaluating their cost using the cost model.

%-- runtime cost models and profiling, unlike SQL planning. 

Once the query has been fully synthesized, \sys schedules the dataflow graph for execution, mapping operators to different nodes while considering their resource capabilities. 
For instance, neural operators may require powerful GPUs (e.g., vision or LLM inference) or remote model invocation (e.g., to GPT-4o), whereas other operators can be executed on CPU platforms. The planner is responsible for identifying optimization opportunities, e.g., based on data and control dependency, execution performance. 
For instance,  independent transformation operators can be applied to the incoming data in parallel. A key difference between the \sys query planner is that it needs a \textit{dynamic} cost model that is updated throughout the query, which further affects how future data processing is performed. 
For instance, a \sys may start with \texttt{Type Unstructured} data, but once it identifies the right structure with LLM processing, the decision could potentially apply to consecutive batches of data until a better structure needs to be extracted.

%% file: tables/table1.tex
\begin{table}[H]
\centering
\caption{The XTP pipeline applied to clinical notes: extracting a relational database, transforming by cleaning and analysis, and projecting a report. Tools: $\mathcal{N}$ (neural), $\mathcal{S}$ (symbolic)}
\label{table:exp-1}
\vspace{-1em}
\tiny
\renewcommand{\arraystretch}{1.2}
\begin{tabular}[t]{|p{0.1cm}|p{1.7cm}|p{5.0cm}|}
\hline
\multicolumn{3}{|p{7.0cm}|}{
\textbf{09/12/2023 – Clinical Notes from Dr. James}

0000h – Patient John Doe (MRN 874521) diagnosed with pneumonia. Prescribed Azithromycin 500mg PO x1 (Brand: Zithromax, GSN 009812, NDC 54569-5821) for bacterial coverage.

0800h – Patient Jane D. (MRN 874522) reported persistent cough. Started on Azithromycin 250mg PO x2 (Brand: Zmax, NDC 60505-3790) for presumed pneumonia.

1100h – John D. returned with worsening symptoms. Physician repeated Azithromycin 500mg PO x1 citing possible resistance, but no culture available.
}
\\
\hline
\vspace{3mm}
\multirow{2}{*}{\rotatebox[origin=c]{90}{\textbf{Extract}}}  &
\texttt{ExtractSchema} \quad \textbf{Tools:} \texttt{GPT-4o}\neural
&
\begin{tabular}{@{}ll@{}}
\textbf{pt}     & \textbf{id}: int [PK], \textbf{name}: string, \textbf{mrn}: string, \textbf{diagnosis}: string \\
\textbf{med}    & \textbf{id}: int [PK], \textbf{name}: string, \textbf{gsn}: string, \textbf{ndc}: string, \textbf{brand}: string, \\
                & \textbf{route}: string, \textbf{dose\_amount}: string, \textbf{dose\_unit}: string \\
\textbf{adm}    & \textbf{id}: int [PK], \textbf{p\_id}: int [FK → pt(id)], \textbf{med\_id}: int [FK → med(id)] \\
\end{tabular}
\\
\cline{2-3}
  &
\begin{tabular}{@{}l@{}}
\texttt{ExtractTriplets} \\
$\rightarrow$ \texttt{ExtractValue} \\
$\rightarrow$ \texttt{ExtractSQL} \\
\textbf{Tools:} \texttt{CoreNLP\symbolic},\\ \texttt{GPT-4o}\neural, \\
\texttt{Python}\symbolic
\end{tabular}
&
\begin{tabular}{@{}l@{}}
("John Doe", "diagnosis", "pneumonia") \\
$\rightarrow$ "1, John Doe, 874521, pneumonia" \\
$\rightarrow$ \texttt{INSERT INTO pt VALUES (1, 'John Doe', '874521', 'pneumonia');}
\end{tabular}
\\
\hline
\vspace{6mm}
\multirow{3}{*}{\rotatebox[origin=c]{90}{\textbf{Transform}}}  &
\begin{tabular}{@{}l@{}}
\texttt{NormalizeBrand} \\
\textbf{Tools:} \texttt{SQL\symbolic}
\end{tabular}
&
\vspace{-1em}
\begin{tabular}{@{}l@{}}
Standardize brand names by replacing them with generic names
\end{tabular}
\hspace{1em}
\begin{tabular}{@{}l@{}}
\textbf{Query:} \\
\texttt{UPDATE med SET name = 'Azithromycin'} \\
\texttt{WHERE brand IN ('Zithromax', 'Zmax');}
\end{tabular}
\\
\cline{2-3}
 &
\begin{tabular}{@{}l@{}}
\texttt{De-Identification} \\
\textbf{Tools:} \texttt{SQL\symbolic}
\end{tabular}
&
\begin{tabular}{@{}l@{}}
Anonymize patient names \\
by removing identifiers from view
\end{tabular}
\hspace{1em}
\begin{tabular}{@{}l@{}}
\textbf{Query:} \\
\texttt{CREATE VIEW anonymized\_patients AS} \\
\texttt{SELECT id, mrn, diagnosis FROM pt;}
\end{tabular}
\\
\cline{2-3}
 &
\begin{tabular}{@{}l@{}}
\texttt{CheckOverPrescription} \\
\textbf{Tools:} \texttt{GPT-4o\neural},\\ \texttt{SQL\symbolic}
\end{tabular}
&
\vspace{-2em}
\begin{tabular}{@{}l@{}}
Detect possible antibiotic overprescription by prompting GPT-4o with database\\ schema and generate SQL query
\end{tabular}
\hspace{1em}
\begin{tabular}{@{}l@{}}
\textbf{Result:} \\
\texttt{id \hspace{1.2em} mrn \hspace{2em} antibiotics} \\
\texttt{1 \hspace{1.2em} 874521 \hspace{3.6em} 3}
\end{tabular}
\\
\hline

\vspace{-1.9cm}\rotatebox[origin=c]{90}{\textbf{Project}}
  &
\multicolumn{2}{p{7.7cm}|}{
  \includegraphics[width=0.8\linewidth]{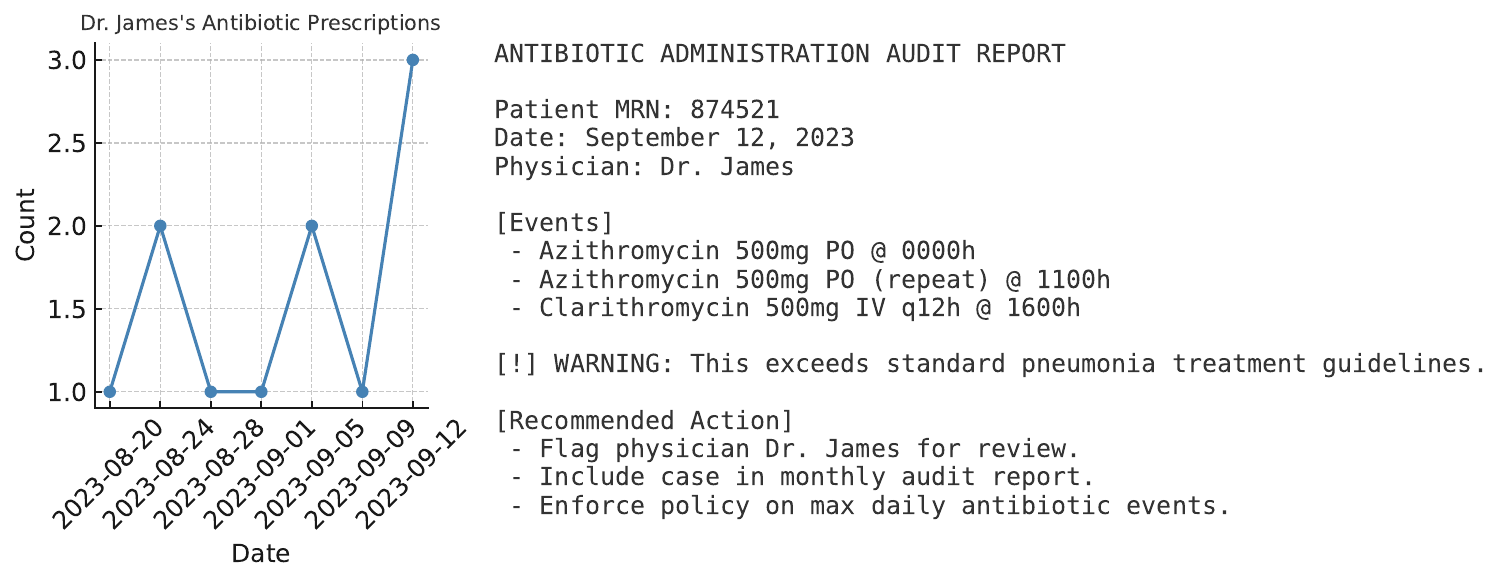}
}
\\
\hline
\end{tabular}

\end{table}

%% file: tables/table2.tex
\vspace{-4mm}
\begin{table}[H]
\caption{The XTP pipeline applied to an electrical circuit: extracting a graph, transforming via validation and redundancy insertion, and projecting a revised diagram. Tools: $\mathcal{N}$ (neural), $\mathcal{S}$ (symbolic)}
\vspace{-2mm}
\label{table:exp-2}
\centering
\tiny
\renewcommand{\arraystretch}{1.2}
\begin{tabular}{|p{0.1cm}|p{2.0cm}|p{5.4cm}|}
\hline
\vspace{-1cm}
\multirow{1}{*}{\rotatebox[origin=c]{90}
{\textbf{Extract}}} &
\includegraphics[width=1\linewidth]{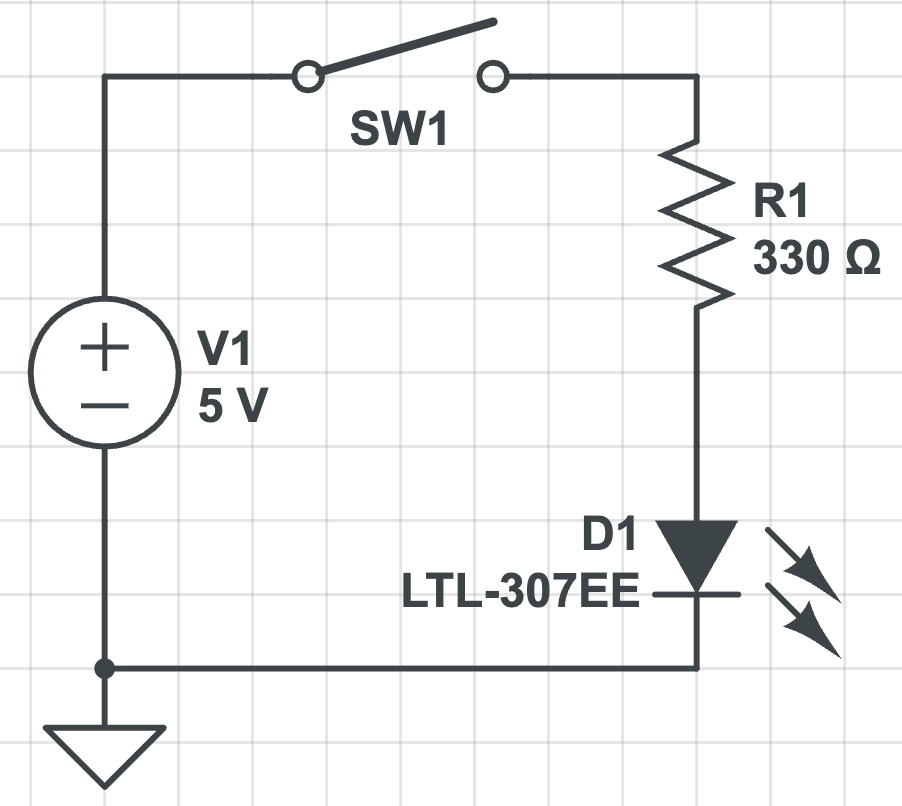}
\begin{tabular}{@{}l@{}}
\texttt{ExtractGraph} \\
\textbf{Tools:} \texttt{GPT-4o vision\neural}
\end{tabular}
&
\vspace{-1.7cm}
\begin{minipage}[t]{\linewidth}
\textbf{Description:} Extract a graph from circuit diagram
\begin{verbatim}
components: [
  {id:1, type:power, label:V+, voltage:12},
  {id:2, type:switch, label:SW},
  {id:3, type:resistor, resistance_ohm:330},
  {id:4, type:led, forward_voltage:2.0},
  {id:5, type:ground, label:GND}
],
connections: [
  {from:V+, to:SW}, {from:SW, to:resistor},
  {from:resistor, to:led}, {from:led, to:GND}]
\end{verbatim}
\end{minipage}
\\
\hline
\vspace{-3mm}
\multirow{3}{*}{\rotatebox[origin=c]{90}{\textbf{Transform}}} &
\begin{tabular}{@{}l@{}}
\texttt{ValidateExtract} \\
\textbf{Tools:} \texttt{Python}\symbolic \\
\textbf{Description:} Encode static \\ rules to validate\\ extracted circuit graph
\end{tabular}
&
\vspace{-0.5cm}
\begin{tabular}[t]{@{}l@{}}
\texttt{Rule: If V\textsubscript{+} - $V_{\text{fwd}}$ > resistor\_limit} $\Rightarrow$ \texttt{Warning} \\[0.4em]
\texttt{For 12V input:} \\
\ \texttt{LED forward voltage} $\approx 2\text{V}$, \texttt{Resistor drop} = $12 - 2 = 10\text{V}$ \\
\texttt{Current} $\approx 10\,\text{V} \div 330\,\Omega \approx 30\text{mA}$ \\
\textbf{\ding{55} Too high for standard LED}
\end{tabular}
\\
\hline
& \texttt{ExtractGraph (re-run)} &
Loops back to extract stage with corrected parameters and re-extracts graph.
\\
\hline
\vspace{0.3cm}
\multirow{3}{*}{\rotatebox[origin=c]{90}{\textbf{Transform}}} &
\begin{tabular}{@{}l@{}}
\texttt{AddRedundancy} \\
\textbf{Tools:} \texttt{GPT-4o}\neural \\
\textbf{Description:} Add a \\redundant branch \\to the circuit
\end{tabular}
&
\vspace{-0.6cm}
\begin{verbatim}
components: [
  ...,
  {id:6, type:resistor, resistance_ohm:330},
  {id:7, type:led, forward_voltage:2.0}
],
connections: [
  ...,
  {from:SW, to:resistor2},
  {from:resistor2, to:led2},
  {from:led2, to:GND}
]
\end{verbatim}
\\
\hline
\vspace{-1.2cm}
\multirow{1}{*}{\rotatebox[origin=c]{90}{\textbf{Project}}} &
\vspace{-1.3cm}
\begin{tabular}{@{}l@{}}
\texttt{ProjectDiagram} \\
\textbf{Tools:} \texttt{GPT-4o image}\neural \\
\textbf{Description:} Generate\\ updated circuit diagram
\end{tabular}

&
\includegraphics[width=0.5\linewidth]{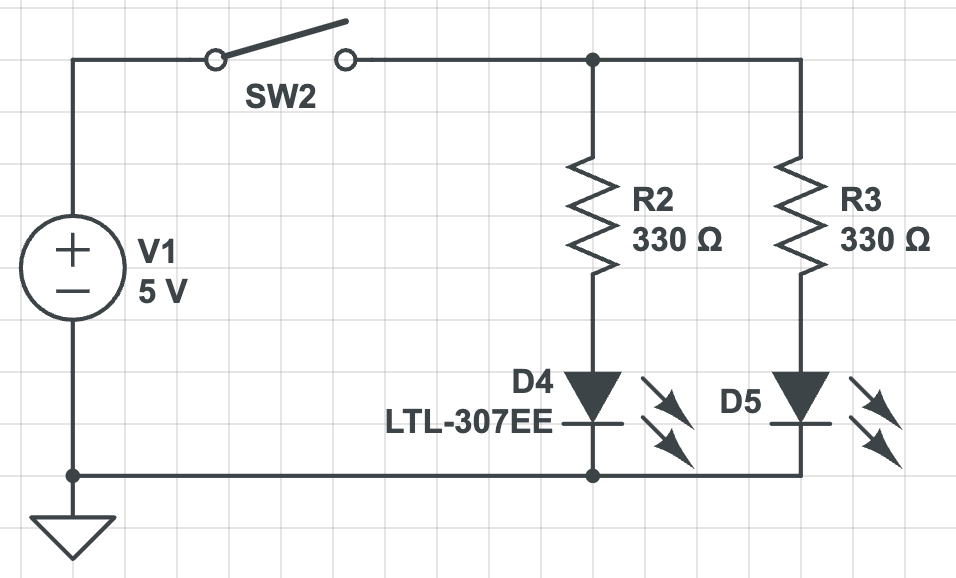}
\\
\hline
\end{tabular}

\end{table}

%% file: sections/5_experiments.tex
\vspace{-5mm}
\section{Experiments}
We present two preliminary experiments, shown in Tables~\ref{table:exp-1} and~\ref{table:exp-2}.\\
\textbf{Experiment 1.} we process clinical notes to generate an anonymized admin report, highlighting overprescription trends. During \textit{extract}, we use an LLM to generate a schema, extract values using a symbolic tool, and align them with the schema using an LLM. SQL statements are then generated symbolically. In \textit{transform}, we clean the data and identify overprescription patterns. Finally, in \textit{project}, we use an LLM to generate the report.\\
\noindent\textbf{Experiment 2.} we analyze a circuit diagram to suggest structural optimizations. In the \textit{extract} phase, a vision model generates the circuit’s topology, followed by symbolic validity checks. If invalid, the system loops back to re-extract. Once valid, \textit{transform} adds a redundant path, and \textit{project} outputs the updated diagram.

These experiments highlight four key aspects of \sys. First, the circuit example demonstrates \textit{dynamic dataflow}, where symbolic validation triggers a loopback to the extract phase. Second, both examples showcase a \textit{mix of neural and symbolic operators}, with symbolic tools verifying and enhancing LLM outputs. Third, the projection phase in the first experiment demonstrates \textit{multimodal output}, where the report includes both a chart and text, highlighting the need for supporting operations over cross-modal data. Fourth, the two tasks differ in complexity: the clinical text example involves diverse value types, requires high accuracy, and thus performs multiple LLM calls in the \textit{extract} phase. In contrast, the simple circuit requires only a single call to a vision model. This highlights the need for a \textit{cost model} to guide query planning by balancing accuracy requirements with latency and token costs.

%% file: sections/related.tex
\vspace{-3.5mm}
\section{Related work}
We organize related work into the following categories.\\
\noindent\textbf{Querying over unstructured data.}
Before LLMs, Systems approached unstructured data using pipelines that relied on rule-based extraction ~\cite{chu2007relational, mansuri}, incurring high overhead and lacking semantic flexibility. More recently, systems like Vizier~\cite{anderson2024design}, Databases Unbound~\cite{sammadden}, and hybrid engines ~\cite{wei2020analyticdb, lin2024towards,lin2025twix} apply LLMs to extract structure, enable semantic search and question answering over raw content. In contrast, \sys goes beyond data access: it performs full computation over extracted structure, enabling read–write pipelines that span extraction, transformation, and projection in multiple modalities.\\
\noindent\textbf{AI techniques in data management.}
Systems like AnDB~\cite{wang2025andbbreakingboundariesainative}, TAG~\cite{biswal2024text2sqlenoughunifyingai}, and LOTUS~\cite{patel2025semanticoperatorsdeclarativemodel} plug LLMs into SQL engines to boost queryability, but stay confined to structured data.
Neurosymbolic pipelines like Galois~\cite{papotti2024large} combine LLMs with rule-based logic, yet rely on fixed execution plans that don't adapt at runtime.
Agent-based tools like Data Agent~\cite{sun2025data} translate natural language instructions into full workflows, but output them as single, opaque scripts—offering no way to inspect, modify, or validate each step.  
In contrast, \sys turns AI components into modular, typed operators that follow a dynamic runtime--where every step can be routed, checked, and adapted on the fly based on data and intent.

%\\\noindent\textbf{Specialized instances.}
\begin{comment}Complementary to the above systems, many domain-specific efforts focus on extracting partial structure from unstructured inputs. Examples include converting text to flat tables~\cite{Aroravldb, sundar2024gtbls}, identifying symbols in diagrams~\cite{sturmer2024diagram, rouabhia2025draw-with-thought, gu2024ai-blueprint}, or performing multimodal information extraction~\cite{sun2024umie}. These approaches often require fine-tuning large models per task, which limits generality, increases cost, and risks catastrophic forgetting~\cite{catastrophic-forgetting}. In contrast, \sys uses a modular and reusable set of neural and symbolic operators orchestrated at runtime, avoiding retraining while enabling broad applicability.
\end{comment}

% \textbf{Projection.} 
% For the reverse direction, generative AI tools can produce unstructured outputs such as text or diagrams. However, there has been little focus on using these tools to systematically close the loop—that is, converting structured data back into unstructured forms. Despite its importance, no existing framework offers a complete, general-purpose solution that supports both directions: from unstructured data to structured representations and back. Bridging this gap remains an open and impactful challenge across domains.

% \textbf{AI and data management.} 
% Find and cite papers. Qizhen Zhang from Toronto has one such paper, but there are multiple groups working on this. 

\vspace{-2mm}
\section{Summary}
\vspace{-1mm}
We describe our vision on enabling extract–transform–project workflows over unstructured inputs. Unlike prior retrieval-focused systems, our envisioned system, \sys, supports full read–write computation using modular neural and symbolic operators, offering a foundation for structured reasoning over unstructured data.
\vspace{-2mm}